# MONG: An extension to Galaxy Clusters

**Louise Rebecca[1,2] • Arun Kenath*[2] • C Sivaram[3]**

[1]Department of Physics, Christ Junior College, Bengaluru, 560029, Karnataka, India

[2]Department of Physics and Electronics, CHRIST (Deemed to be University), Bengaluru, 560029, Karnataka, India

[3]Indian Institute of Astrophysics, Bengaluru, 560034, Karnataka, India

E-mail addresses: louise.rheanna@res.christuniversity.in (Louise Rebecca); kenath.arun@christuniversity.in (Arun Kenath); sivaram@iiap.res.in (C Sivaram)

**Abstract**: The presence of dark matter, though well established by indirect evidence, is yet to be observed directly. Various dark matter detection experiments running for several years have yielded no positive results. In view of these negative results, we had earlier proposed alternate models by postulating a minimum gravitational field strength (minimum curvature) and a minimum acceleration. These postulates led to the modified Newtonian dynamics and modified Newtonian gravity (MONG). The observed flat rotation curves of galaxies were also accounted for through these postulates. Here we extend these postulates to galaxy clusters and model the dynamical velocity-distance curve for a typical cluster such as the Virgo cluster. The radial velocities of galaxies in the Virgo cluster are also obtained through this model. Observations show an inconsistency in the Hubble flow at a mean cluster distance of 17 Mpc, which is expected in regions of high matter density. This decrease in velocity is predicted by our model of modified gravity (MONG). The radial velocity versus distance relation for galaxies in the Virgo cluster obtained using MONG is in agreement with observations.

---

*Corresponding author

# 1. Introduction

The nature of dark matter (DM) and dark energy (DE), which constitutes more than 95% of the energy density of the Universe, is still uncertain. Though there is indirect evidence for the presence of DM in the Universe, its nature and distribution remain unknown for the most part. Experiments for the detection of DM particles have been running for years with no positive results so far; only lower and lower limits for their fluxes (for different DM particle masses) are set [1]. This perhaps necessitates alternate models to dark matter [2].

Dark matter plays an important role in the observed dynamics of large-scale structures. One of the key evidence of its presence is the rotation curves of galaxies [3]. The manner in which the velocities vary with radius gives the distribution of mass in a galaxy. For a solid disk, the velocity would vary linearly with the radius. If the mass is concentrated at the center of the galaxy, the velocity decreases with the square root of the radius (Keplerian case). The rotation curves of most galaxies have two regions. The inner region is linear, implying that velocities increase linearly with the radius. The outer region, i.e. the one corresponding to distances away from the center is flat suggesting that the velocities remain constant at the outer regions until the visible edge of galaxies. A flat rotation curve thus implies that the mass is still increasing with radius in the outer regions. This led to the conclusion that there is a halo of invisible mass (dark matter) surrounding galaxies.

The modification of Newtonian Dynamics (MOND) proposed by Milgrom [4] as an alternative to dark matter accounts for observed flat rotation curves of galaxies. But this theory involved an ad hoc introduction of a fundamental acceleration, $a_0 \approx 10^{-8} cms^{-2}$. As the acceleration approaches $a_0$, the Newtonian law gets modified and the field strength takes the form,

$$a = \frac{(GMa_0)^{1/2}}{r} \qquad (1)$$

where, $a$ is the acceleration, $r$ is the radial distance and $M$ is the central mass. Equation (1) gives a constant velocity $\left(v_c = (GMa_0)^{1/4}\right)$ at the galactic outskirts, thus accounting for the flat rotation curves without invoking the need for DM.

These results can also be arrived at by considering a minimum acceleration corresponding to minimum gravitational field strength (at the outskirts of galaxies and galaxy clusters) given by [5],

$$a_{min} = \frac{GM}{r_{max}^2} \qquad (2)$$

Here, $a_{min} \approx 10^{-8} cms^{-2}$ plays the same role as the MOND acceleration $a_0$, below which the Newtonian dynamics gets modified, giving rise to the observed galactic dynamics. By postulating a minimum possible acceleration, we eliminate the ad-hoc nature of the MOND acceleration ($a_0$). This postulate gives rise to a maximum radius, $r_{max}$ (through equation (2)) corresponding to the minimum acceleration, i.e., it is the maximum possible size to which a large scale structure of mass $M$ can grow. Indeed, in an earlier work, equation (2) was used to apply constraints on the maximal sizes of galaxies and galaxy clusters [6].

Further substituting for $r_{max}$ (from Equation (2)) in the usual expression for the velocity of stars in the galactic outskirts, we have

$$v_c^2 = \frac{GM}{r_{max}} = \frac{GM}{\sqrt{\frac{GM}{a_{min}}}} \tag{3}$$

From equation (3), we get

$$v_c = (GMa_{min})^{1/4} \tag{4}$$

This is same as the constant velocity in the galactic outskirts as proposed by MOND with the minimum acceleration, $a_{min}$, here playing the role of MOND acceleration, $a_0$. Equation (4) is also consistent with the Tully-Fisher relation (expression). Hence, this postulate of minimal acceleration (field strength) leads to the same results as MOND (like for example as seen from equation (4)).

## 2. Modification of Newtonian Gravity (MONG)

In earlier papers [5-7] we have shown that the flat rotation curves can alternatively be explained by considering Modifications to Newtonian Gravity (MONG) by adding an additional gravitational self-energy density term, along with the dark energy cosmological constant term (acting at larger distances) to the Poisson's equation. Here we extend these considerations (i.e., MONG) to the observed velocity-distance relations for large galaxy clusters such as the Virgo cluster. The gravitational self-energy density also contributes to the gravitational field along with matter density $\rho$. With this modification, we have:

$$\nabla^2 \phi = 4\pi G \rho + K(\nabla \phi)^2 + \Lambda c^2 \tag{5}$$

where, $\phi \left(= \frac{GM}{r}\right)$ is the usual gravitational potential and the constant $K \approx \left(\frac{G^2}{c^2}\right)$. $K(\nabla \phi)^2$ is the gravitational self-energy density [6]. The matter density is considered to be uniform in the regions closer to the galactic centre up to a distance $r_o$ after which it falls off with distance as,

$$\rho = \rho_o \left(\frac{r_o}{r}\right)^2 \tag{6}$$

For regions within $r_o$ – where matter density dominates and is a constant – the equation (5) is the usual Poisson's equation (i.e., $\nabla^2 \phi = 4\pi G \rho$) with the solution $\phi = \frac{GM}{r}$. For a constant density ($\rho_o$) and mass $M = \frac{4}{3}\pi r_0{}^3 \rho_o$, the velocity is given by

$$v = \left(\frac{4}{3}\pi G \rho_o\right)^{1/2} r \qquad (7)$$

Thus, giving a velocity that varies linearly with distance up to a distance $r_0$. At larger distances from the centre of the galaxy, $> r_0$, the matter density falls off as given by equation (6) and the gravitational self-energy dominates. The equation (5) now takes the form (neglecting the constant $\Lambda$ (cosmological constant) term),

$$\nabla^2 \phi = 4\pi G \rho_o \left(\frac{r_o}{r}\right)^2 + K(\nabla\phi)^2 \qquad (8)$$

The solution of this equation yields:

$$\phi = (Q + K') \ln \frac{r}{r_o} \qquad (9)$$

where $Q = 4\pi G \rho_o r_o{}^2$ and $K' \approx \frac{GM}{r_o}$ are constants.

The regions near the outskirts of a large structure are subjected to minimum gravitational field strength with minimum acceleration given by equation (2) [6,8]. This minimum acceleration ($a_{min} \approx 10^{-8} cms^{-2}$) [8] is of the same order as that of the MOND acceleration [4,9,10]. Equation (9) gives a force (per unit mass) of the form,

$$F = \frac{K''}{r} \qquad (10)$$

where $K'' = (GMa_{min})^{1/2}$ is also a constant.

The balance of the centripetal force and gravitational force then gives,

$$\frac{V^2}{r} = \frac{K''}{r} \qquad (11)$$

Equation (11) gives the relation $v^2 = K''$

Thus, implying the independence of velocity ($v$) on distance ($r$), as indicated by the flat rotation curve.

For larger distances from the center of the galaxy/cluster (i.e. $\gg r_o$) the matter density is very small and the dark energy dominates. Then, equation (5) now takes the form

$$\nabla^2 \phi = K(\nabla\phi)^2 + \Lambda c^2 \qquad (12)$$

where, $\Lambda$ is the constant cosmological constant term (DE term).

The solution of the above equation yields

$$\phi = A \ln \sec Br \qquad (13)$$

where $A \approx (GMa_{min})^{1/2}$ and $B \approx \sqrt{\Lambda}c(GMa_{min})^{-1/4}$ are constants. This gives a force (per unit mass) of the form

$$F = \nabla \phi = B' tan Br \qquad (14)$$

where $B' \approx \sqrt{\Lambda}c(GMa_{min})^{1/4}$ is also a constant.

On expanding equation (14) we get,

$$F = \nabla \phi = B'\left(Br + \frac{1}{3}(Br)^3 + \cdots\right) \qquad (15)$$

The outskirts of the large-scale structure are subjected to a minimum acceleration of $10^{-8} cms^{-2}$. For the Virgo cluster with baryonic mass $M \approx 4 \times 10^{47} g$, the constants take the values of: $B \approx 10^{-27} cm^{-1}$ and $B' \approx 10^{-9} cms^{-2}$. Hence, at larger distances (outskirts of the cluster) the higher order terms of equation (15) become negligible. The dependence of the constant $B'$ on the mass $M$ of galaxy cluster is small, even an increase in mass by two orders would yield a value nearly same as above. This gives a force (per unit mass) of the form,

$$F = \Lambda c^2 r \qquad (16)$$

The distance from the centre of the structure at which the dark energy term starts to dominate can be estimated from equation (16). The constraints on this distance (after which DE dominates) can be set by incorporating the balance of gravitational self-energy and the repulsive dark energy as the boundary condition, i.e.:

$$\frac{(GMa_{min})^{\frac{1}{2}}}{r} = \Lambda c^2 r' \qquad (17)$$

This gives,

$$r' = \sqrt{\frac{(GMa_{min})^{1/2}}{\Lambda c^2}} \qquad (18)$$

For the Virgo cluster, with a luminous mass, $M \approx 4 \times 10^{47} g$, and with the observed cosmological constant term of $\Lambda = 10^{-56} cm^{-2}$, this distance ($r'$) turns out to be $4 \times 10^{25} cm$ ($14 Mpc$). The dark energy (DE) dominated region can be observed in galaxy clusters rather than individual galaxies as this region is at a much greater distance (at least by two orders) than the typical size of galaxies.

The balance of the centripetal force and (modified) gravitational force now gives,

$$\frac{V^2}{r} = \Lambda c^2 r \qquad (19)$$

The above equation implies a linear variation of velocity with distance. Thus, this model predicts an increase in velocity with distance for regions farther away from the inner regions of the large-scale structures (galaxy clusters). The velocity variation with distance for a typical galaxy cluster like the Virgo cluster is plotted in Figure 1, the transition (at $14 Mpc$) to linear increase in velocity with distance is marked.

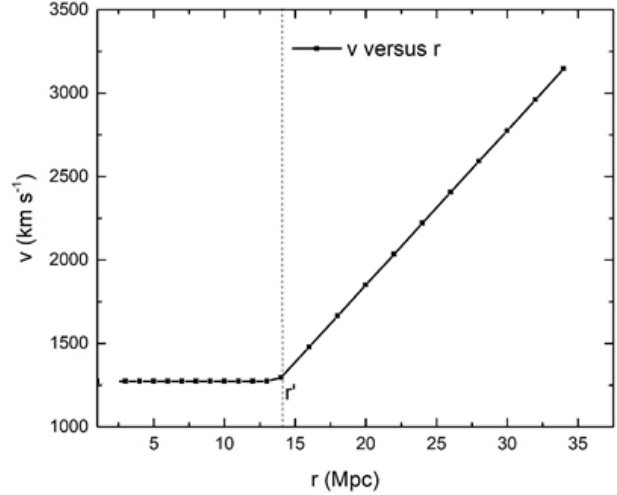

**Figure 1** Rotational velocity versus distance relation for galaxies in the Virgo cluster.

The solution of equation (5) is thus used to predict different regimes of interest in a typical galaxy cluster such as the Virgo cluster. Where matter density dominates and is a constant ($\rho_0$), the velocity varies linearly with distance. For $r > r_0$ and within $r'$ the gravitational self-energy term dominates giving a flat region (constant velocity) to the rotation curve. Beyond $r'$ the curve shifts back to linear behaviour as the dark energy term begins to dominate. The predictions made by our model can be supported by future observations on galaxy cluster dynamics.

## 3. Infall of galaxies in the Virgo cluster

The results of MONG can also be used to make predictions on the radial velocities of galaxies in the Virgo cluster. The radial velocity estimates from MONG are in good agreement with observations of the infall of galaxies in the Virgo cluster [13]. It is well known that the velocity-distance relation in the immediate vicinity of the Local Group (LG) must deviate from the usual Hubble flow as a result of the deceleration caused by the LG mass [11,12]. By estimating the radius of the zero-velocity surface, the total mass of the LG can be determined independently from mass estimates based on orbital motion [13]. The radius of the zero-velocity surface is estimated observationally [13] by analysing the velocity-distance relation in Virgocentric coordinates using a sample of 1792 galaxies whose distances have been estimated using different methods such as the tip of the Red Giant Branch, the Cepheid luminosity, the luminosity of type Ia supernovae, the surface brightness fluctuation method and the Tully–Fisher relation. The radius of the zero-velocity surface is thus estimated to lie at a mean cluster distance of $17 Mpc$. This seems consistent with relevant observations [13].

We also arrive at this non-linearity in the Hubble flow and predict the radial velocities of galaxies around the Virgo-cluster. Around the immediate vicinity of the LG, the local gravitational effect is compensated by the background expansion so that the net gravitational field strength is almost zero. This region is thereby subjected to a small net field strength with the acceleration given by,

$$a = \frac{dv}{dt} = \frac{(GMa_0)^{1/2}}{R} \qquad (20)$$

where, $v$ is the radial velocity at a distance $R$. Integrating equation (20) gives a velocity of the form,

$$v = v_0 \sqrt{2 \ln\left(\frac{R}{R_c}\right)} \qquad (21)$$

$v_0 \left(\approx (GMa_0)^{1/4}\right)$ is the velocity corresponding to the minimal acceleration. For the Virgo cluster with $M \approx 4 \times 10^{47} g$, considering only the baryonic mass, this velocity turns out to be around $1278\ kms^{-1}$ (as indicated in fig 1). $R_c$ is the radius corresponding to the velocity $v_0$. From observational data $R_c$ is found to be around $17 Mpc$ [13]. With these estimates, we plot the radial velocity-distance relation for the Virgo cluster in Figure 2 and compare it to that obtained from observations.

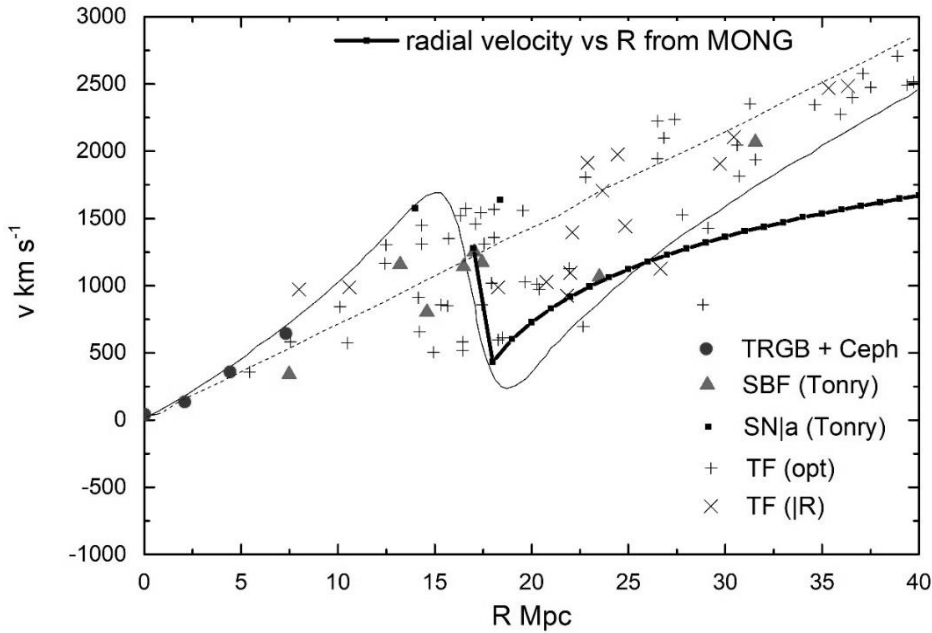

**Figure 2** Radial velocity versus distance relation for galaxies in the Virgo cluster. Our model (thick curve) compared to the perturbation in Hubble flow from observations [13] (s shaped

thin curve) overlaid on observational points. The inclined dotted line traces the Hubble flow with $H_0 = 72 \, kms^{-1} Mpc^{-1}$. TRGB is the tip of the red giant branch and SBF is surface brightness fluctuations.

## 4. Conclusion

The addition of the dark energy density and a gravitational self-energy density to the Poisson's equation leads to the Modification of Newtonian Gravity and with the introduction of a minimum acceleration in Newtonian dynamics the flat rotation curves of galaxies and galaxy clusters is modeled without invoking DM. The region close to the center of the galaxy clusters, where the matter density dominates and is uniform i.e., $< r_0$, gives a linear increase in velocity with distance.

For distances $r > r_0$, the matter density falls off with distance and the gravitational self-energy term dominates giving a velocity that is independent of distance implying a flat region in the rotation curve, thereby accounting for dark matter (DM). The regions at the outskirts of the galaxy cluster (where the dark energy dominates) are subjected to a minimum gravitational field strength and a minimum acceleration ($a_0 \approx 10^{-8} cms^{-2}$) which was introduced in an earlier work to avoid an ad hoc fundamental acceleration accounting for dark matter. As dark energy begins to dominate, the velocity is found to increase linearly with distance. In the case of the Virgo cluster, the dark energy is found to dominate from a distance of $14 Mpc$ (Fig. 1) where the curve shifts from being flat to linear. Future observational estimates of galaxy cluster dynamics could support our approach. We have also extended our work to the radial velocities of galaxies in the Virgo cluster and show that the inconsistency in the Hubble flow obtained by observations is predicted by the present model of modified gravity.

We also note that these DM effects can be, in principle and at least partially, be explained through the framework of extended gravity (without requiring DM) [14, 15].